\documentclass{article}
\usepackage{amsmath, amsthm}
\usepackage{amsfonts}
\usepackage{amssymb}
\usepackage{fullpage}
\usepackage[noblocks]{authblk}
\usepackage{hyperref}
\usepackage{fixmath}
\usepackage{tikz}
\usepackage{subfigure}
\usetikzlibrary{arrows,shapes}
\usepackage{xifthen}
\usepackage{mathptmx}

\newcommand{\set}[1]{{\{#1\}}}
\newcommand{\Zplus}{\ensuremath{\mathbb{Z}^{+}}}
\newcommand{\pair}[1]{\ensuremath{#1}} 
\newcommand{\seg}[1]{\ensuremath{\overline{#1}}}
\newcommand{\ray}[1]{\ensuremath{\overrightarrow{#1}}}
\newcommand{\lin}[1]{\ensuremath{\overleftrightarrow{#1}}} 
\newcommand{\comp}[1]{{\ensuremath{\overline{#1}}}} 
\newcommand{\nice}[1]{\ensuremath{#1}-\emph{straight}}
\newcommand{\polite}[1]{\ensuremath{#1}-\emph{convex}}
\newcommand{\flanked}[1]{\ensuremath{#1}-\emph{reflex}}
\newcommand{\dangerous}[1]{\ensuremath{#1}-\emph{fragmented}}
\newcommand{\dang}{fragmented}
\newcommand{\ch}[1]{\ensuremath{conv(#1)}}
\newcommand{\tri}{\Delta}	

\def\abs#1{\left| #1 \right|}

\newtheorem{theorem}{Theorem}[section]

\newtheorem{lemma}[theorem]{Lemma}

\theoremstyle{definition}

\theoremstyle{remark}

\tikzstyle{every node}=[draw,circle,fill=black, minimum size=3pt, inner sep=0pt]
\tikzstyle{blue node}=[draw,circle,fill=cyan!30!white, minimum size=3pt, inner sep=0pt]
\tikzstyle{red node}=[draw,circle,fill=red!75!black, minimum size=3pt, inner sep=0pt]
                            
\tikzstyle{edge} = [draw,thin,-,black]
\tikzstyle{nonedge} =  [draw,-,black,dashed]
\tikzstyle{thick nonedge} =  [draw,thick,-,black,dashed]
\tikzstyle{thick edge} = [draw,thick,black]

\tikzstyle{obedge} = [draw,thick,-,fill=green]
\tikzstyle{obedge2} = [draw,line width=2pt,-,green]
\tikzstyle{smallob} = [draw,rectangle,fill=green, minimum size = 10pt]
\tikzstyle{vsmallob} = [draw,rectangle,fill=green, minimum size = 5pt]

\tikzstyle{carve} = [draw,yellow!20,fill=yellow!20]
\tikzstyle{carve2} = [draw,green!20,fill=green!20]
\tikzstyle{carvewhite} = [draw,white,fill=white]

\author{J{\'{a}}nos Pach}
\affil{\'Ecole Polytechnique F\'ed\'erale de Lausanne\authorcr
Station 8, CH-1015 Lausanne, Switzerland\authorcr
Email address available at
\href{http://cims.nyu.edu/\textasciitilde pach/address.html}{cims.nyu.edu/\textasciitilde pach}\authorcr
\ }

\author{Den{\.{i}}z Sar{\i}{\"{o}}z}
\affil{The Graduate Center of The City University of New York\authorcr
365 Fifth Avenue, New York, NY 10016 USA\authorcr
Email address available at \href{http://sarioz.com}{sarioz.com}}

\title{Small $(2,s)$-colorable graphs without 1-obstacle representations\footnote{Research supported by NSA grant 47149-00 01, NSF grant CCF-08-30272, Swiss National Science Foundation grant 200021-125287/1, and by the Bernoulli Center at EPFL.}}

\begin{document}
\maketitle
\begin{abstract}
An obstacle representation of a graph $G$ is a 
set of points on the plane 
together with a set of polygonal obstacles that determine
a visibility graph isomorphic to $G$.
The obstacle number of $G$
is the minimum number of obstacles
over all obstacle representations of $G$.

Alpert, Koch, and Laison \cite{AKL09} gave a 12-vertex bipartite graph
and proved that its obstacle number is \emph{two}.
We show that a 10-vertex induced subgraph of this graph has obstacle number \emph{two}.

Alpert et al. \cite{AKL09} also constructed very large graphs with vertex set consisting of a clique and an independent set
in order to show that obstacle number is an unbounded parameter.
We specify a 70-vertex graph with vertex set consisting of a clique and an independent set, 
and prove that it has obstacle number greater than \emph{one}.

This is an ancillary document to our article in press \cite{PS10_GraphsCombin}.
We conclude by showing that a 10-vertex graph with vertex set consisting of two cliques
has obstacle number greater than \emph{one}, improving on a result therein.
\end{abstract}

\thispagestyle{empty} 
\section{Introduction}
\label{sec:intro}
Consider a finite set $P$ of points on the plane,
and a set of closed polygonal obstacles whose vertices
together with the points in $P$ are in general position, that is, no three of them are collinear.
The corresponding visibility graph has $P$ as its vertex set,
two points $p, q \in P$ having an edge between them 
if and only if the line segment $pq$ does not meet any obstacles.
Visibility graphs are extensively studied and used in computational geometry and robot motion planning; 
see \cite{BKOS00, G07, OR97, O99, Ur00}.

Relatively recently, Alpert, Koch, and Laison \cite{AKL09} introduced an interesting new parameter of graphs, closely related to visibility graphs.
Given a graph $G$, we say that a set of points and a set of polygonal obstacles as above constitute an \emph{obstacle representation} of $G$, if the corresponding visibility graph is isomorphic to $G$. A representation with $h$ obstacles is called an $h$-obstacle representation. The smallest number of obstacles in an obstacle representation of $G$ is called the \emph{obstacle number} of $G$.

A graph is called $(r,s)$-colorable \cite{BT97_hered}
if its vertex set can be partitioned into $r$ sets, $s$ of which are cliques and $r-s$ of which are independent sets.
For instance, $(2,0)$-colorable graph is simply a bipartite graph, a 
$(2,1)$-colorable graph is a split graph \cite{FH77_splitdef, TCh79},
and a $(2,2)$-colorable graph has bipartite complement.

In our paper in press \cite{PS10_GraphsCombin},
we employed extremal graph theoretic methods to show that for every constant $h$,
the number of graphs on $n$ vertices with obstacle number at most $h$ is $2^{o(n^2)}$,
based on the graphs $G_1$, $G_2$, and $G_3$ with the properties stated in the following.
In this ancillary note to that paper, we accomplish three tasks.
We show that a particular 10-vertex 
$(2,0)$-colorable (i.e., bipartite) graph $G'_{1}$ has
obstacle number greater than \emph{one}.
This improves upon the 12-vertex bipartite graph $G_1$ in \cite{AKL09}, 
and settles a conjecture therein.
We also show that a particular 70-vertex $(2,1)$-colorable graph $G'_{2}$ has obstacle number greater than \emph{one},
improving on the 
$\big(92379 + {92379 \choose 6}\big)$-vertex 
graph implied by a construction in \cite{AKL09}.
In \cite{PS10_GraphsCombin}, we had given a $(2,2)$-colorable
20-vertex graph $G_3$, and proved that it has obstacle number greater than \emph{one}.  
We finally show that a related $(2,2)$-colorable 10-vertex graph $G'_3$ also has obstacle number greater than \emph{one}.

\section{A 10-vertex bipartite graph without a \mbox{1-obstacle} representation}
\label{sec:Kstarmns}
Given a graph,
we refer to a distinct pair of vertices of the graph that does not define an edge of the graph as a non-edge.
In every drawing of a simple finite graph there is bound to be a unique unbounded face,
referred to as the outside face.
A 1-obstacle representation in which the obstacle lies on the outside face is called an 
outside obstacle representation, and such an obstacle is called an outside obstacle.

In \cite{AKL09},
$K^{*}_{m,n}$ 
has been defined as
the graph obtained from the complete bipartite graph
$K_{m,n}$ by removing a maximum matching.
There, it was shown that every $K^{*}_{m,n}$ graph admits
a 2-obstacle representation:
The two independent sets are placed within disjoint half-planes,
such that the non-edges in the removed matching meet at a single point
so that a single non-outside obstacle is sufficient to meet them,
while the non-edges within the independent sets meet the outside face
so that an outside obstacle is sufficient to meet them.
The authors also gave a strong hint for obtaining an outside obstacle representation
of $K^{*}_{4,n}$ for every $n$ 
by providing an easily generalizable outside obstacle representation for $K^{*}_{4,5}$.
Furthermore, they proved that 
$G_1 := K^{*}_{5,7}$ does not admit a \mbox{1-obstacle} representation.

\begin{figure*}[htp]
\begin{center}
\begin{tikzpicture}[scale=0.8]

\path[draw = black] (1,-2) rectangle (9,5);

\path (3, 0) node[red node] (b1) [label=below:$r_1$]{};
\path (4.1, -0.4) node[red node] (b2) [label=below:$r_2$]{};
\path (5, -0.5) node[red node] (b3) [label=below:$r_3$]{};
\path (6.1, -0.4) node[red node] (b4) [label=below:$r_4$]{};
\path (7, 0) node[red node] (b5) [label=below:$r_5$]{};

\path (7, 3) node[blue node] (r1) [label=above:$b_1$]{};
\path (6, 3.4) node[blue node] (r2) [label=above:$b_2$]{};
\path (4.94, 3.5) node[blue node] (r3) [label=above:$b_3$]{};
\path (4, 3.4) node[blue node] (r4) [label=above:$b_4$]{};
\path (3.01, 3) node[blue node] (r5) [label=above:$b_5$]{};

\newcommand{\maxb}{5}
\newcommand{\maxr}{5}

\foreach \i in {1,...,\maxb}
{
	\foreach \j in {1,...,\maxb}
	{
		\ifthenelse{\i > \j} {}
		{ 
			\draw[nonedge] (b\i) -- (b\j);
		}
	}
}
\foreach \i in {1,...,\maxr}
{
	\foreach \j in {1,...,\maxr}
	{
		\ifthenelse{\i > \j} {}
		{ 
			\draw[nonedge] (r\i) -- (r\j);
		}
	}
}
\foreach \i in {1,..., \maxb}
{
	\foreach \j in {1,..., \maxr}
	{
		\ifthenelse{\equal{\i}{\j}} {%
		\draw[nonedge] (b\i) -- (r\j);
		} { 
		\draw[edge] (b\i) -- (r\j);
		}
	}
}
\path node [smallob] (ob1) at (5, 1.5)  {};

\draw[obedge2] (3.5, 0.2)   
				 -- ++ (0.7, -1.8) -- ++ (0.4, 1.3) 
				 -- ++ (0.4, -1.2) -- ++ (0.6, 1.2)
				 -- ++ (0.6, -1) -- ++ (0.4, 1.4)
				 -- ++ (0.3, -1) -- ++ (0.5, 0) %
				 -- ++ (0, 5.2) -- ++ (-0.4, 0) %
				 -- ++ (-0.5, -1.5) -- ++ (-0.5, 1.5)
				 -- ++ (-0.5, -1) -- ++ (-0.7, 1.2)
				 -- ++ (-0.3, -1.3) -- ++ (-0.6, 1.2)
				 -- ++ (-0.5, -1.6);
\end{tikzpicture}
\end{center}
\caption{A 2-obstacle representation of $G'_1$, i.e., $K^{*}_{5,5}$.}
\label{fig:KStar5comma5optrep}
\end{figure*}

We dedicate the rest of this section to proving their following conjecture.

\begin{theorem}
\label{KStar5Comma5HasObsNum2}
$G'_1 := K^{*}_{5,5}$, 
the graph obtained from $K_{5,5}$ by removing a perfect matching,
has obstacle number 2.
\end{theorem}

\begin{proof}
%
To able to refer to individual vertices of $K^{*}_{5,5}$,
let $V(K^{*}_{5,5}) = B \uplus R$
such that
$B = \set{b_1, b_2, b_3, b_4, b_5}$ (the set of light blue vertices)
and $R = \set{r_1, r_2, r_3, r_4, r_5}$ (the set of dark red vertices)
are independent sets 
and there is an edge from a blue vertex $b_i$ 
to a red vertex $r_j$ 
if and only if $i \neq j$.

Before we proceed, we borrow some definitions and two facts 
from \cite{AKL09}.

Given points $a$, $b$, $c$ in the plane we say $a$ sees $b$ to the left of $c$
(equivalently, sees $c$ to the right of $b$) if the points $a$, $b$, and $c$
appear in clockwise order.
If a point $a$ is outside the convex hull of some set $S$ of points, the relation
``$a$ sees to the left of'' is transitive on $S$, hence is a total ordering of $S$,
called the $a$-sight ordering of $S$.

We paraphrase Lemma 3 of \cite{AKL09} in the following way.

\begin{lemma}
\label{K23byAKL}
If a graph having $K_{2,3}$ as an induced subgraph
has a 1-obstacle representation,
then in such a representation
the two parts (independent sets) of the induced $K_{2,3}$ are linearly separable.
Moreover, for each part $S$, 
every vertex in the other part induces the same sight ordering of $S$.
\end{lemma}

Lastly, we paraphrase a fact used in the original proof of Lemma~\ref{K23byAKL}.

\begin{lemma}
\label{bowtie}
In a 1-obstacle representation of $K^{*}_{5,5}$,
every vertex subset $S$ consisting of 2 red vertices and 2 blue vertices with 4 distinct subscripts
(the necessary and sufficient condition for a $K_{2,2}$ to be induced)
is in convex position, 
with both color classes appearing contiguously 
around the convex hull of $S$.
Hence the drawing induced on $S$
(i.e., the drawing of every induced $K_{2,2}$ in $K^{*}_{5,5}$) 
is self-intersecting, a \emph{bowtie}.
\end{lemma}

We now give and prove a new lemma, one of many
to help prune the space of vertex arrangements
potentially amenable to 1-obstacle representations of 
$K^{*}_{5,5}$.

For any three points $p, q, r$, we denote by
$\angle p q r$ the union of the rays
\ray{qp} and \ray{qr}.
We denote by \ch{P} for the convex hull of a point set $P$.
\begin{lemma}
\label{TheObstacleIsOutside}
Every 1-obstacle representation of 
$K^{*}_{5,5}$
is an outside obstacle representation.
\end{lemma}

\begin{proof}
Assume that we are 
given a 1-obstacle representation of $K^{*}_{5,5}$
that is not an outside obstacle representation.
At least three vertices are on the convex hull boundary of the vertices
by the general position assumption.
Every pair of vertices appearing consecutively around the
convex hull boundary must constitute an edge,
otherwise an outside obstacle would be required to block it.
Then without loss of generality $b_1, r_2, b_3$ appear consecutively on the bounding polygon.
All other vertices including $r_4$ are inside $\ch{\angle b_1 r_2 b_3}$.
Hence the drawing of the $K_{2,2}$ induced on
$\set{b_1, r_2, b_3, r_4}$
is not a bowtie,
which contradicts Lemma~\ref{bowtie}.
\end{proof}
%

\begin{lemma}
\label{SightOrderingsExistForIncidentVertices}
In every 1-obstacle representation of 
$K^{*}_{5,5}$,
every vertex $v$ is linearly separable from the set $S$ of its neighbors,
defining a $v$-sight ordering on $S$.
\end{lemma}

\begin{proof}
Assume for contradiction that we are given
a 1-obstacle representation of $K_{5,5}$
in which some vertex, without loss of generality, $b_1$,
is not linearly separable from the set of its neighbors.
Then $b_1$ is in the convex hull of $\set{r_2, r_3, r_4, r_5}$.
By the general position assumption,
a triangulation of $\set{r_2, r_3, r_4, r_5}$
will reveal that $b_1$ is inside some triangle with red vertices,
Without loss of generality, $\tri r_3 r_4 r_5$.
Then by the general position assumption,
the ray $\ray{b_1 b_2}$ meets an interior point of some edge of this triangle,
Without loss of generality, $\seg{r_4 r_5}$.
This implies that the drawing of 
$K_{2,2}$ induced on $\set{b_1, r_4, b_2, r_5}$ is not a bowtie,
which contradicts Lemma~\ref{bowtie}.
\end{proof}

In a graph drawing or obstacle representation,
we say that a polygon is \emph{solid} 
if it is a subset of the drawing: 
if every point on it is a vertex or on an edge.

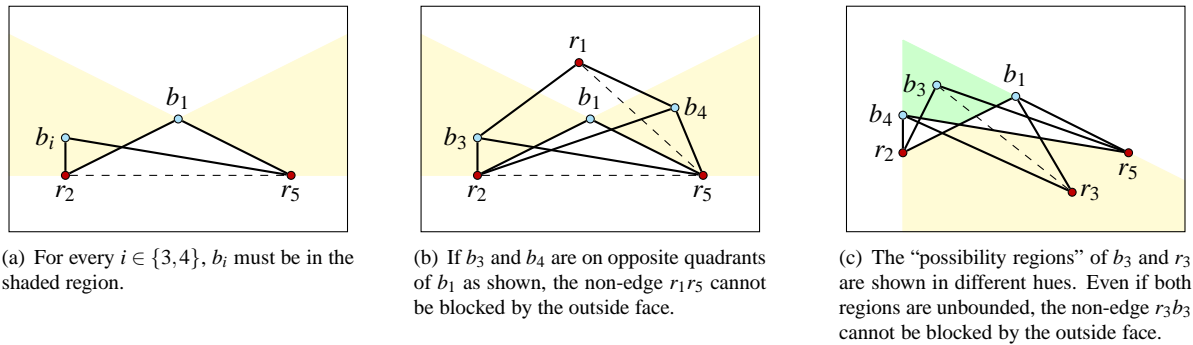
\begin{figure*}[htp]
\newcommand{\figymin}{-2}
\newcommand{\figymax}{2}
\begin{center}
\subfigure[For every $i \in \set{3,4}$,
$b_i$ must be in the shaded region.]{
\begin{tikzpicture}[scale=0.75]
\newcommand{\figxmin}{-3}
\newcommand{\figxmax}{3}
\draw[carve] (\figxmin, 1.5) -- (0, 0) -- (-2, -1) -- (\figxmin, -1) -- (\figxmin, 1.5);
\draw[carve] (\figxmax, 1.5) -- (0, 0) -- (2, -1) -- (\figxmax, -1) -- (\figxmax, 1.5);

\path (0, 0) node[blue node] (b1) [label=above:$b_1$]{};
\path (-2, -1) node[red node] (r2) [label=below:$r_2$]{};
\path (2, -1) node[red node] (r5) [label=below:$r_5$]{};
\path (-2, -0.333333333333333) 		node[blue node] 	(bi)	[label=left:$b_i$]{};

\draw[thick edge] (r2) -- (b1) -- (r5);
\draw[thick edge] (r5) -- (bi) -- (r2);
\draw[nonedge] (r2) -- (r5);

\path[draw = black] (\figxmin, \figymin) rectangle (\figxmax, \figymax);
\end{tikzpicture}
\label{fig:goodQuadrants}
} 
\qquad
\subfigure[If $b_3$ and $b_4$ are on opposite quadrants of $b_1$ as shown,
the non-edge $\pair{r_1 r_5}$ cannot be blocked by the outside face.]{
\begin{tikzpicture}[scale=0.75]
\newcommand{\figxmin}{-3}
\newcommand{\figxmax}{3}
\draw[carve] (\figxmin, 1.5) -- (0, 0) -- (-2, -1) -- (\figxmin, -1) -- (\figxmin, 1.5);
\draw[carve] (\figxmax, 1.5) -- (0, 0) -- (2, -1) -- (\figxmax, -1) -- (\figxmax, 1.5);

\path (0, 0) 		node[blue node] 	(b1)	[label=above:$b_1$]{};
\path (-2, -1) 		node[red node]		(r2)	[label=below:$r_2$]{};
\path (2, -1) 		node[red node] 	(r5)	[label=below:$r_5$]{};
\path (-2, -0.333333333333333) 		node[blue node] 	(b3)	[label=left:$b_3$]{};
\path (1.5, 0.2) 		node[blue node] 	(b4)	[label=right:$b_4$]{};

\path (-0.2, 1) 		node[red node] 	(r1)	[label=above:$r_1$]{};

\draw[thick edge] (r2) -- (b1) -- (r5);
\draw[thick edge] (r5) -- (b3) -- (r2);
\draw[thick edge] (r5) -- (b4) -- (r2);
\draw[nonedge] (r2) -- (r5);

\draw[thick edge] (b3) -- (r1) -- (b4);
\draw[nonedge] (r1) -- (r5);

\path[draw = black] (\figxmin, \figymin) rectangle (\figxmax, \figymax);
\end{tikzpicture}
\label{fig:oppositeQuadrantsBad}
} 
\qquad
\renewcommand{\figymax}{1.6}
\renewcommand{\figymin}{-2.4}
\subfigure[The ``possibility regions'' of $b_3$ and $r_3$ are shown in different hues.
Even if both regions are unbounded, 
the non-edge $\pair{r_3 b_3}$ cannot be blocked by the outside face.]{
\begin{tikzpicture}[scale=0.75]
\newcommand{\figxmin}{-3}
\newcommand{\figxmax}{3}

\draw[carve] (-2, \figymin) -- (-2, -1) -- (-1, -0.5) -- (2, -1) -- (\figxmax, -1.5) -- (\figxmax, \figymin) --  (-2, \figymin);
\draw[carve2] (-2, 1) -- (0, 0) -- (-1, -0.5) -- (-2, -0.333333333333333) -- (-2,1);

\path (0, 0) 		node[blue node] 	(b1)	[label=above:$b_1$]{};
\path (-2, -1) 		node[red node]		(r2)	[label=left:$r_2$]{};
\path (2, -1) 		node[red node] 	(r5)	[label=below:$r_5$]{};
\path (-2, -0.333333333333333) 		node[blue node] 	(b4)	[label=left:$b_4$]{};
\path (-1.4, 0.2)	node[blue node] 	(b3)	[label=left:$b_3$]{};
\path (1, -1.7) node[red node] (r3) [label=right:$r_3$]{};

\draw[thick edge] (r2) -- (b1) -- (r5);
\draw[thick edge] (r5) -- (b3) -- (r2);
\draw[thick edge] (r5) -- (b4) -- (r2);
\draw[thick edge] (b4) -- (r3) -- (b1);
\draw[nonedge] (r3) -- (b3);

\path[draw = black] (\figxmin, \figymin) rectangle (\figxmax, \figymax);
\end{tikzpicture}
\label{fig:possibilityRegions}
} 
\end{center}
\caption{Subfigures \subref{fig:goodQuadrants}, \subref{fig:oppositeQuadrantsBad}, and
\subref{fig:possibilityRegions}
respectively accompany the second, third, and last paragraphs
in the proof of \ref{EveryVertexSeparableFromOtherPart}.
Some edges and non-edges are omitted for clarity, 
as they often will be in subsequent figures.} 
\label{fig:EveryVertexSeparableFromOtherPart}
\end{figure*}

\begin{lemma} 
\label{EveryVertexSeparableFromOtherPart}
In every 1-obstacle representation of 
$K^{*}_{5,5}$,
every vertex in $R$ (respectively, $B$) is linearly separable from $B$ (respectively, $R$).
\end{lemma}

\begin{proof}
We will show that in every 1-obstacle representation of 
$K^{*}_{5,5}$, each blue vertex is linearly separable from $R$.
The analogous statement about each red vertex and $B$
can be proved symmetrically.

Assume for contradiction that we are given
a 1-obstacle representation of $K^{*}_{5,5}$ 
in which some blue vertex is in \ch{R}.
Without loss of generality, 
$b_1 \in \ch{R}$.
By Lemma~\ref{SightOrderingsExistForIncidentVertices},
$b_1$ is linearly separable from $\set{r_2, r_3, r_4, r_5}$.
Without loss of generality, $\lin{r_2 r_5}$ is a horizontal line with $r_2$ to the left of $r_5$
such that $b_1$ is above $\lin{r_2 r_5}$ and
$\pair{r_3 r_4}$ is inside $\ch{\angle r_2 b_1 r_5}$.
Call the four open regions delineated by the lines
$\lin{r_2 b_1}$ and $\lin{r_5 b_1}$ the
left, right, upper, and lower $b_1$-quadrants.
For each $i \in \set{3,4}$, 
since the drawing of $K_{2,2}$ induced on
$\set{r_2, b_i, r_5, b_1}$ must be a bowtie by Lemma~\ref{bowtie},
$b_i$ is 
above $\lin{r_2 r_5}$ and
in either the left or the right $b_1$-quadrant.  
(See Fig.~\ref{fig:goodQuadrants}.)

Without loss of generality, $b_3$ is in the left $b_1$-quadrant.
Assume for contradiction that $b_4$ is in the right $b_1$-quadrant.
Since $b_1 \in \ch{R}$,
$r_1$ is in the upper \mbox{$b_1$-quadrant}.
Then $b_3, r_1, b_4$ are 
respectively in the left, upper, and right $b_1$-quadrants,
and $r_5$ is on the boundary of the right and lower $b_1$-quadrants.
This implies that the drawing of $K_{2,2}$ induced on $\set{b_3, r_1, b_4, r_5}$
is non-self-intersecting, not a bowtie.
By Lemma~\ref{bowtie}, 
this means $b_4$ is in the left $b_1$-quadrant along with $b_3$.
(See Fig.~\ref{fig:oppositeQuadrantsBad}.)

Notice that $K_{2,3}$ is induced on $\{r_2, r_5, b_1, b_3, b_4\}$.
Then by Lemma~\ref{K23byAKL}, 
$b_3$ and $b_4$ are above $\lin{r_2 r_5}$
along with $b_1$,
and the \mbox{$r_2$- and $r_5$-sight orderings} of
$\set{b_1, b_3, b_4}$ are the same, 
with $b_1$ appearing rightmost.
Without loss of generality, $r_2$ and $r_5$ see 
$b_4$ to the left of $b_3$.
Hence, $b_3$ is inside $\ch{\angle b_4 r_2 b_1}$
in addition to being inside $\ch{\angle b_4 r_5 b_1}$.
By the same token,
since $b_1$ sees $r_3$ to be between $r_2$ and $r_5$, so does $b_4$.
Hence,
$r_3$ is inside $\ch{\angle r_2 b_4 r_5}$,
in addition to being inside $\ch{\angle r_2 b_1 r_5}$.
These conditions ensure that 
$\ch{\angle r_2 b_3 r_5}$ and $\ch{\angle b_4 r_3 b_1}$
meet to give a convex quadrilateral region with solid boundary
that has $\seg{b_3 r_3}$ as a diagonal.
This implies that the non-edge $\pair{b_3 r_3}$ 
is not blocked by the outside face, 
in contradiction to Lemma~\ref{TheObstacleIsOutside}.
(See Fig.~\ref{fig:possibilityRegions}.)
\end{proof}

Denote by $K^{-}_{3,3}$ 
the graph obtained by removing an edge from $K_{3,3}$.
Note that our proof of Lemma~\ref{EveryVertexSeparableFromOtherPart} 
relies on showing that the assumptions lead to a
drawing of $K_{2,2}$ forbidden by Lemma~\ref{bowtie},
or to a forbidden drawing of $K^{-}_{3,3}$ like the one shown in
Fig.~\ref{fig:possibilityRegions}.

\begin{lemma}
\label{RnBCHDisjoint} 
In every 1-obstacle representation of 
$K^{*}_{5,5}$,
the convex hulls of $R$ and $B$ are disjoint, hence, 
there is a line separating $R$ from $B$.
\end{lemma}

\begin{proof}
Assume for contradiction that we are given a 
1-obstacle representation of 
$K^{*}_{5,5}$
in which $\ch{R} \cap \ch{B} \neq \emptyset$.
Let $X$ denote $\ch{R} \cap \ch{B}$.
But by Lemma~\ref{EveryVertexSeparableFromOtherPart},
$(R \cup B) \cap X = \emptyset$.
This means that $X$ is a $2k$-gonal shape ($2 \leq k \leq 5$) separating
\ch{B} and \ch{R} into $k$ pieces each, alternating around it.

If $k \geq 3$,
Without loss of generality, $r_1 b_{i_2} r_2 b_{i_1} r_3 b_{i_3}$
is a counterclockwise enumeration of some convex hexagon $H$.
Take \lin{r_2 r_3} as horizontal.
Without loss of generality, $b_4$ is below \lin{r_2 r_3}.
By Lemma \ref{K23byAKL},
$b_1$ and $b_5$ are also below \lin{r_2 r_3}.
This means that $\set{b_{i_2}, b_{i_3}} = \set{b_{2}, b_{3}}$.
If $i_2 = 3$, then $H$ is solid and has
the non-edge \pair{b_2 b_3} as an internal diagonal,
which therefore requires an internal obstacle, 
contradicting \ref{TheObstacleIsOutside}.
Otherwise, $i_2 = 2$ and
\seg{r_2 b_3} meets \seg{b_2 r_3} at some point $q$,
so the solid convex quadrilateral
$r_1 b_2 q b_3$
has
\pair{b_2 b_3} 
as an internal diagonal,
which once again requires an internal obstacle, 
contradicting \ref{TheObstacleIsOutside}.

Therefore,
$X$ separates \ch{R} and \ch{B} into 2 pieces each.
Denote by $R_1$ and $R_2$ the subsets of $R$ induced by this partition,
and define $B_1$ and $B_2$ similarly.
Without loss of generality, $|R_1| \in \set{1, 2}$ and $|B_1| \in \set{1, 2}$.
Now we will show that $\abs{R_1} = \abs{B_1} = 1$.

Assume otherwise for contradiction.
Without loss of generality, $R_1 = \set{r_1, r_2}$ and $R_2 = \set{r_3, r_4, r_5}$.
By Lemma~\ref{K23byAKL},
$\seg{r_1 r_3}$ is linearly separable from $\tri b_2 b_4 b_5$.
Clearly, $\lin{r_1 r_3}$ 
separates $B_1$ from $B_2$.
This implies that $\set{b_2, b_4, b_5} \subseteq B_2$.
Similarly, $\seg{r_2 r_4}$ is linearly separable from  
$\tri b_1 b_3 b_5$,
which implies $\set{b_1, b_3, b_5} \subseteq B_2$.
But then we have $|B_2| = 5$, a contradiction.

Without loss of generality, let $R_1= \set{r_1}$.  
To see that this forces $B_1 = \set{b_1}$,
assume for contradiction that (without loss of generality) $B_1 = \set{b_2}$.
Then $\seg{r_1 r_3}$ meets $\seg{b_2 b_4}$,
contradicting Lemma~\ref{bowtie}.

Without loss of generality, the sets
$R_1, B_1, R_2, B_2$ appear clockwise around $X$,
in this order.
Notice that every red vertex in $R_2$ sees $b_1$ rightmost in $B$.
Without loss of generality, let the \mbox{$b_1$-sight ordering} of $R$ be
$r_5, r_4, r_3, r_2, r_1$.
To highlight the resemblance to the proof of 
Lemma~\ref{EveryVertexSeparableFromOtherPart},
take the line $\lin{r_2 r_5}$ to be horizontal with $r_2$ to the left of $r_5$.

Since $K_{2,3}$ is induced on $\set{b_1, b_3, b_4, r_2, r_5}$,
by Lemma~\ref{K23byAKL}
the $r_2$- and $r_5$-sight orderings of
$\set{b_4, b_3, b_1}$ are the same.
Since
$r_2$ and $r_5$ are in $R_2$, they
see $b_1$ as the rightmost blue vertex
and without loss of generality they 
see $b_4$ to the left of $b_3$.
Thus we have exactly the same conditions 
as those used in the last paragraph of the proof of
Lemma~\ref{EveryVertexSeparableFromOtherPart}
to conclude that 
$\seg{b_3 r_3}$
is an interior diagonal of a solid quadrilateral,
hence an outside obstacle is insufficient in this case too.

Therefore, in a 1-obstacle representation of 
$K^{*}_{5,5}$,
\ch{B} and \ch{R} are disjoint.
\end{proof}


Armed with the knowledge that every 1-obstacle representation of
$K^*_{5,5}$ is an outside obstacle representation and 
requires $R$ and $B$ to be linearly separable,
assume for contradiction that we are given a drawing of 
$K^{*}_{5,5}$ that admits a 1-obstacle representation.
We will argue that such a drawing
necessarily contains a
drawing of $K_{2,2}$ requiring more than one obstacle
or a drawing of $K^{-}_{3,3}$ 
requiring more than one obstacle.
We justify the existence of such a forbidden configuration 
by using an algorithm
that removes vertices from the drawing until casually inspecting 
the convex hull boundary of the vertices must
reveal the existence of such a configuration.

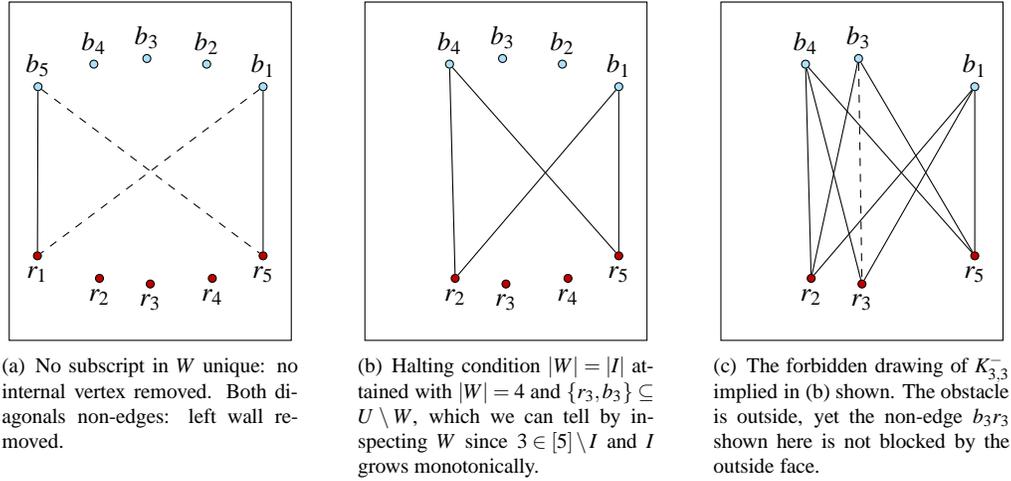
\begin{figure*}[htp]

\begin{center}
\subfigure[No subscript in $W$ unique: no internal vertex removed.
Both diagonals non-edges: left wall removed.]{
\begin{tikzpicture}[scale=0.75]

\path[draw = black] (2.5,-1.5) rectangle (7.5,4.5);

\path (3, 0) node[red node] (r1) [label=below:$r_1$]{};
\path (4.1, -0.4) node[red node] (r2) [label=below:$r_2$]{};
\path (5, -0.5) node[red node] (r3) [label=below:$r_3$]{};
\path (6.1, -0.4) node[red node] (r4) [label=below:$r_4$]{};
\path (7, 0) node[red node] (r5) [label=below:$r_5$]{};

\path (7, 3) node[blue node] (b1) [label=above:$b_1$]{};
\path (6, 3.4) node[blue node] (b2) [label=above:$b_2$]{};
\path (4.94, 3.5) node[blue node] (b3) [label=above:$b_3$]{};
\path (4, 3.4) node[blue node] (b4) [label=above:$b_4$]{};
\path (3.01, 3) node[blue node] (b5) [label=above:$b_5$]{};

\draw[nonedge] (b1) -- (r1);
\draw[nonedge] (b5) -- (r5);
\draw[edge] (b1) -- (r5);
\draw[edge] (b5) -- (r1);
\end{tikzpicture}
} 
\qquad
\subfigure[Halting condition $\abs{W} = \abs{I}$ attained with $\abs{W} = 4$ and $\set{r_3, b_3} \subseteq U \setminus W$,
which we can tell by inspecting $W$ since \mbox{$3 \in [5] \setminus I$}
and $I$ grows monotonically.]{
\begin{tikzpicture}[scale=0.75]

\path[draw = black] (2.5,-1.5) rectangle (7.5,4.5);

\path (4.1, -0.4) node[red node] (r2) [label=below:$r_2$]{};
\path (5, -0.5) node[red node] (r3) [label=below:$r_3$]{};
\path (6.1, -0.4) node[red node] (r4) [label=below:$r_4$]{};
\path (7, 0) node[red node] (r5) [label=below:$r_5$]{};

\path (7, 3) node[blue node] (b1) [label=above:$b_1$]{};
\path (6, 3.4) node[blue node] (b2) [label=above:$b_2$]{};
\path (4.94, 3.5) node[blue node] (b3) [label=above:$b_3$]{};
\path (4, 3.4) node[blue node] (b4) [label=above:$b_4$]{};

\draw[edge] (b1) -- (r5);
\draw[edge] (b1) -- (r2);
\draw[edge] (b4) -- (r5);
\draw[edge] (b4) -- (r2);

\end{tikzpicture}
\label{fig:haltingWith4WallVertices}
} 
\qquad
\subfigure[The forbidden drawing of $K^{-}_{3,3}$ implied in
\subref{fig:haltingWith4WallVertices} shown.
The obstacle is outside, yet
the non-edge $\pair{b_3 r_3}$ shown here is not blocked by the outside face.]{
\begin{tikzpicture}[scale=0.75]

\path[draw = black] (2.5,-1.5) rectangle (7.5,4.5);

\path (4.1, -0.4) node[red node] (r2) [label=below:$r_2$]{};
\path (5, -0.5) node[red node] (r3) [label=below:$r_3$]{};
\path (7, 0) node[red node] (r5) [label=below:$r_5$]{};

\path (7, 3) node[blue node] (b1) [label=above:$b_1$]{};
\path (4.94, 3.5) node[blue node] (b3) [label=above:$b_3$]{};
\path (4, 3.4) node[blue node] (b4) [label=above:$b_4$]{};

\draw[edge] (b1) -- (r5);
\draw[edge] (b1) -- (r2);
\draw[edge] (b4) -- (r5);
\draw[edge] (b4) -- (r2);
\draw[edge] (r2) -- (b3) -- (r5);
\draw[edge] (b1) -- (r3) -- (b4);
\draw[nonedge] (b3) -- (r3);
\end{tikzpicture}
} 

\end{center}
\caption{A run of the algorithm in the proof of 
Lemma \ref{KStar5Comma5HasObsNum2}.
Notice that the initial state
features a placement of $V(K^{*}_{5,5})$ in which $R$ is linearly separable from $B$
and below $B$ as required.
In all but the last subfigure, only the dichromatic pairs induced on $W$ are shown.
}
\label{fig:algo1-KStar5comma5}
\end{figure*}

Now we give some terminology needed to describe the algorithm.
By Lemma \ref{RnBCHDisjoint} the convex hulls of $B$ and $R$ are disjoint,
so let the $x$-axis separate $B$ and $R$
with $B$ above it and $R$ below it.
Then for $U \subseteq V(K^{*}_{5,5})$
s.t.
$\abs{U \cap R} \geq 3$ and $\abs{U \cap B} \geq 3$,
in a clockwise walk around the boundary of \ch{U}
there is a unique clockwise-ordered pair of consecutive vertices 
of the form $(r_i, b_j)$ and 
a unique clockwise-ordered pair of consecutive vertices 
of the form $(b_k, r_\ell)$
by the general position assumption.
Call $\set{r_i, b_j}$ the left wall of $U$ and denote it by 
$w_{\mathrm{left}} = w_{\mathrm{left}}(U)$.
Likewise, call $\set{b_k, r_\ell}$ the right wall of $U$ and denote it by 
$w_{\mathrm{right}} = w_{\mathrm{right}}(U)$.
Let $W = W(U) = \set{r_i, b_j, b_k, r_\ell}$, the wall vertices of $U$.
The assumptions on $U$
imply
$3 \leq \abs{W} \leq 4$.
Denote by $I = I(U)$ the set of subscripts occurring in $W(U)$.
Then $2 \leq \abs{I} \leq 4$.
Observe that $\abs{W} - \abs{I}$ is the number of dichromatic non-edges of $K^{*}_{5,5}$
induced on $W$.

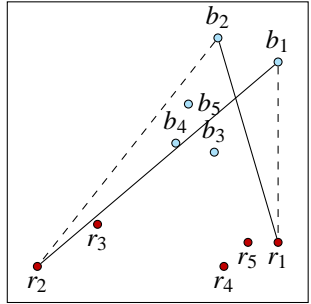
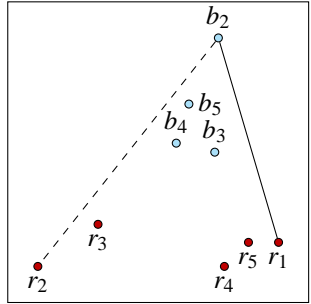
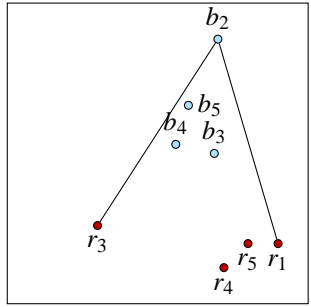
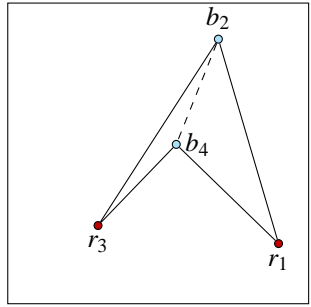
\begin{figure*}[htp]

\begin{center}
\subfigure[
$\abs{W} = 4$ and $w_\mathrm{right}$ is a non-edge: Removing $b_1$
will not evict $r_1$ from the right wall, so $b_1$ is removed.]{
\begin{tikzpicture}[scale=0.8]
\path[draw = black] (2.5,-1) rectangle (7.5,4);
\path (3, -0.4) node[red node] (r2) [label=below:$r_2$]{};
\path (4, 0.3) node[red node] (r3) [label=below:$r_3$]{};
\path (6.1, -0.4) node[red node] (r4) [label=below:$r_4$]{};
\path (6.5, 0) node[red node] (r5) [label=below:$r_5$]{};
\path (7, 0) node[red node] (r1) [label=below:$r_1$]{};
\path (7, 3) node[blue node] (b1) [label=above:$b_1$]{};
\path (6, 3.4) node[blue node] (b2) [label=above:$b_2$]{};
\path (5.94, 1.5) node[blue node] (b3) [label=above:$b_3$]{};
\path (5.3, 1.65) node[blue node] (b4) [label=above:$b_4$]{};
\path (5.51, 2.3) node[blue node] (b5) [label=right:$b_5$]{};
\draw[nonedge] (b1) -- (r1);
\draw[nonedge] (b2) -- (r2);
\draw[edge] (b1) -- (r2);
\draw[edge] (b2) -- (r1);
\end{tikzpicture}
} 
\qquad
\subfigure[
\mbox{$\abs{W} = 3$} and $w_\mathrm{left}$ is the unique dichromatic non-edge induced on $W$.
The vertex $b_2$ is kept since it is in both walls, while its twin $r_2$ is removed.
]{
\begin{tikzpicture}[scale=0.8]
\path[draw = black] (2.5,-1) rectangle (7.5,4);
\path (3, -0.4) node[red node] (r2) [label=below:$r_2$]{};
\path (4, 0.3) node[red node] (r3) [label=below:$r_3$]{};
\path (6.1, -0.4) node[red node] (r4) [label=below:$r_4$]{};
\path (6.5, 0) node[red node] (r5) [label=below:$r_5$]{};
\path (7, 0) node[red node] (r1) [label=below:$r_1$]{};
\path (6, 3.4) node[blue node] (b2) [label=above:$b_2$]{};
\path (5.94, 1.5) node[blue node] (b3) [label=above:$b_3$]{};
\path (5.3, 1.65) node[blue node] (b4) [label=above:$b_4$]{};
\path (5.51, 2.3) node[blue node] (b5) [label=right:$b_5$]{};
\draw[nonedge] (b2) -- (r2);
\draw[edge] (b2) -- (r1);
\end{tikzpicture}
} 
\end{center}
\begin{center}
\subfigure[Halting condition \mbox{$\abs{W} = \abs{I}$} attained with 
\mbox{$\abs{W} = 3$}
and
\mbox{$b_4 \in U \setminus W$},
which we can tell by inspecting $W$ since \mbox{$4 \in [5] \setminus I$} and $I$ grows monotonically.
]{
\begin{tikzpicture}[scale=0.8]
\path[draw = black] (2.5,-1) rectangle (7.5,4);
\path (4, 0.3) node[red node] (r3) [label=below:$r_3$]{};
\path (6.1, -0.4) node[red node] (r4) [label=below:$r_4$]{};
\path (6.5, 0) node[red node] (r5) [label=below:$r_5$]{};
\path (7, 0) node[red node] (r1) [label=below:$r_1$]{};
\path (6, 3.4) node[blue node] (b2) [label=above:$b_2$]{};
\path (5.94, 1.5) node[blue node] (b3) [label=above:$b_3$]{};
\path (5.3, 1.65) node[blue node] (b4) [label=above:$b_4$]{};
\path (5.51, 2.3) node[blue node] (b5) [label=right:$b_5$]{};
\draw[edge] (b2) -- (r3);
\draw[edge] (b2) -- (r1);
\end{tikzpicture}
\label{fig:haltingWith3WallVertices}
} 
\qquad
\subfigure[The forbidden drawing of $K_{2,2}$ implied in 
\subref{fig:haltingWith3WallVertices} shown.
The obstacle is outside, yet
the non-edge \pair{b_2 b_4} is not blocked by the outside face.]{
\begin{tikzpicture}[scale=0.8]
\path[draw = black] (2.5,-1) rectangle (7.5,4);
\path (4, 0.3) node[red node] (r3) [label=below:$r_3$]{};
\path (7, 0) node[red node] (r1) [label=below:$r_1$]{};
\path (6, 3.4) node[blue node] (b2) [label=above:$b_2$]{};
\path (5.3, 1.65) node[blue node] (b4) [label=right:$b_4$]{};
\draw[edge] (b2) -- (r3) -- (b4) -- (r1) -- (b2);
\draw[nonedge] (b2) -- (b4);
\end{tikzpicture}
} 
\end{center}

\caption{Another run of the algorithm in the proof of
Lemma \ref{KStar5Comma5HasObsNum2},
illustrating configurations distinct from those shown in 
Fig.~\ref{fig:algo1-KStar5comma5}.
}
\label{fig:algo2-KStar5comma5}
\end{figure*}

Here is the algorithm sketch.
Initialize $U := V(K^{*}_{5,5})$.
The halting condition is  $\abs{W} = \abs{I}$,
i.e., that every vertex in $W$ has a distinct subscript.
Repeat the following until the halting condition arises.
For every $i \in [5]$, we say that $r_i$ and $b_i$ are twins.
For every vertex with a unique subscript in $W$, 
remove its twin from $U$ (unless it has already been removed).
Remove at least one vertex in $W$ with a twin also in $W$,
the specifics to be described later.

A vertex $v$ is removed from $U$ only if its twin $\comp{v}$ is in $W$,
and removing $v$ will cause $\comp{v}$ to be locked in $W$ for the rest of the algorithm execution,
due to the careful way in which we remove a wall vertex.
Assuming that this claim holds, $I$ grows monotonically.
This means 
$\set{r_j, b_j} \subseteq U \setminus W$
for every $j \in [5] \setminus {I}$.
Let us call the vertices in $U \setminus W$ the interior vertices of $U$,
and a pair $\set{r_j, b_j} \subseteq U \setminus W$ an interior non-edge of $U$.
Since $\abs{I} \leq \abs{W} \leq 4$ and $I$ grows monotonically, $U$ always has some interior non-edge.
Furthermore, at most two vertices from each color class are ever removed, ensuring the
propagation of the precondition
$\abs{U \cap R} \geq 3$ and $\abs{U \cap B} \geq 3$ and proper termination.
We now show why the halting condition implies a forbidden configuration.
The halting condition $\abs{W} = \abs{I}$ arises in two cases:

\begin{enumerate}

\item $\abs{W} = 3$.
Without loss of generality, $W = \set{b_1, r_2, b_3}$.
Then $r_4$ is an interior vertex of $U$ by the monotonicity of $I$.
We will show that the copy of $K_{2,2}$ induced on
$\set{b_1, r_2, b_3, r_4}$ gives a contradiction.
$r_4$ is inside $\ch{\angle b_1 r_2 b_3}$,
hence the drawing of the $K_{2,2}$ induced on
$\set{b_1, r_2, b_3, r_4}$ is no bowtie,
which by Lemma~\ref{bowtie} yields a contradiction.

\item $\abs{W} = 4$.
Without loss of generality, $w_{\mathrm{left}}  = \set{r_1, b_2}$ and $w_{\mathrm{right}}  = \set{b_3, r_4}$.
Then $\set{b_5, r_5}$ is an interior non-edge of $U$ by the monotonicity of $I$.
We will show that the copy of $K^{-}_{3,3}$ 
induced on $\set{r_1, r_4, r_5, b_2, b_3, b_5}$ gives a contradiction.
Notice that $K_{2,3}$ is induced on 
$\set{r_1, r_4, b_2, b_3, b_5}$
and on
$\set{r_1, r_4, r_5, b_2, b_3}$.
Clearly $b_2$ is the leftmost vertex in the $r_1$-sight ordering of $\set{b_2, b_3, b_5}$,
$r_1$ is the rightmost vertex in the $b_2$-sight ordering of $\set{r_1, r_4, r_5}$,
$b_3$ is the rightmost vertex in the $r_4$-sight ordering of $\set{b_2, b_3, b_5}$,
and 
$r_4$ is the leftmost vertex in the $b_3$-sight ordering of $\set{r_1, r_4, r_5}$.
By applying Lemma~\ref{K23byAKL} to the aforementioned two vertex sets 
on which $K_{2,3}$ is induced,
we obtain that 
$b_2$ and $b_3$ both see $r_5$ between $r_1$ and $r_4$,
and that
$r_1$ and $r_4$ both see $b_5$ between $b_2$ and $b_3$.
These conditions are sufficient to ensure that 
$\pair{b_5 r_5}$ 
is an internal diagonal of a solid quadrilateral and hence
cannot be blocked by the outside face,
contradicting Lemma~\ref{TheObstacleIsOutside}.

\end{enumerate}

Now we describe how to remove wall vertices in a way that guarantees the ``locking''
described above, and hence the monotonicity of $I$.
Note that removing a vertex does not affect a wall that it is not in.

If $|W| =4$, $w_{\mathrm{left}} = \set{r_i, b_j}$, and $w_{\mathrm{right}} = \set{b_k, r_\ell}$, then
we call $\set{r_i, b_k}$ and $\set{b_j, r_\ell}$ the diagonals of $U$.
If both diagonals of $U$ are non-edges,
remove from $U$ both vertices in $w_{\mathrm{left}}$.
If a single diagonal of $U$ is a non-edge,
then without loss of generality, $w_{\mathrm{left}} = \set{r_1, b_2}$ and 
$w_{\mathrm{right}} = \set{b_1, r_3}$.
In this case, proceed to the next iteration by removing $b_1$ from $U$.
Now we argue why this ensures that $r_3$ gets locked in the right wall.
For every $i \in \set{4,5}$,
$K_{2,2}$ is induced on $\set{b_1, r_3, b_2, r_i}$,
hence by Lemma~\ref{bowtie}, $r_i \notin \mathrm{int} \angle b_2 r_3 b_1$.
Recalling that $r_2$ has already been removed,
the next counterclockwise vertex after $r_3$ on the resulting convex hull boundary
after removing $b_1$ will still be blue.
Therefore, $r_3$ remains in the right wall.

If some wall is a non-edge, then without loss of generality, 
$w_{\mathrm{right}} = \set{b_1, r_1}$.
If $\abs{W} = 3$, 
Without loss of generality, $w_{\mathrm{left}} = \set{r_2, b_1}$.
Remove $r_1$ from $U$, so that $r_2$ and $b_1$ will be locked in $w_{\mathrm{left}}$.
If $\abs{W} = 4$, pick the vertex to remove from $w_{\mathrm{right}}$ in the following way.
If $r_{1} \in w_{\mathrm{right}} (U \setminus \set{b_1})$ then
remove $b_1$, otherwise remove $r_1$.
To show why this simple action
guarantees that the twin of the removed vertex `stays' in the right wall,
we need to justify that if $r_{1} \notin w_{\mathrm{right}} (U \setminus \set{b_1})$
then $b_1 \in w_{\mathrm{right}} (U \setminus \set{r_1})$.

By hypothesis, 
$w_{\mathrm{right}} (U \setminus \set{b_1}) = \set{b', r'}$
where $r' \in R \setminus \set{r_1}$.
First we must explain why $b_1$ is the unique blue vertex in $U$ 
to the right of the line $\lin{r'b'}$.
By the definition of right wall,
no vertex in $U \setminus \set{b_1}$ is to the right of the line $\lin{r'b'}$.
But if $b_1$ were also to the left of the line $\lin{r'b'}$,
then $r'$ together with $b'$ would constitute the right wall of $U$,
contradicting $\set{b_1, r_1} = w_{\mathrm{right}}(U)$.
Therefore, $b_1$ is the unique blue vertex to the right of $\lin{r'b'}$.
Initialize a dynamic line $L$ to $\lin{r'b'}$.
Rotate $L$ clockwise around \ch{U \setminus \set{b_1, r_1}} 
until it becomes horizontal, allowing it to sweep the entire
portion of the half-plane above the $x$-axis to the right of $\lin{r'b'}$.
Clearly, $b_1$ is the unique blue vertex of $U$ swept by $L$.
Denote by $\hat{r}$ the other vertex of $U\setminus{\set{r_1}}$ on 
$L$ at the precise moment when $b_1$ is swept by $L$, 
which is unique by the general position assumption.
It is possible that ${\hat{r}} = r'$.
No vertex of $U \setminus{\set{r_1}}$ is to the right of the line
$\lin{\hat{r} b_1}$.  
Therefore,
$\set{\hat{r}, b_1} = w_{\mathrm{right}}(U \setminus \set{r_1})$.

This completes an informal and yet complete specification of the algorithm
that shows that every 1-obstacle representation of $K^*_{5,5}$ has a forbidden configuration of vertices
resulting in a contradiction.
Therefore, the obstacle number of $G'_1$, i.e., $K^*_{5,5}$, is greater than \emph{one}.
This implies that the obstacle number of $G'_1$ is \emph{two},
per its obstacle representation in Fig.~\ref{fig:KStar5comma5optrep}.

\end{proof}

\section{A 70-vertex $(2,1)$-colorable graph without a 1-obstacle representation}
\label{sec:CENotAllAtMost1}
\begin{theorem}
\label{CE6HasNoOneObstacleRep}
The $(2,1)$-colorable graph 
$G'_2 := CE(6)$,
consisting of a clique of 6 blue vertices
and an independent set of 64 red vertices 
each of which has a distinct set of neighbors,
has obstacle number greater than \emph{one}.
\end{theorem}
\begin{center}
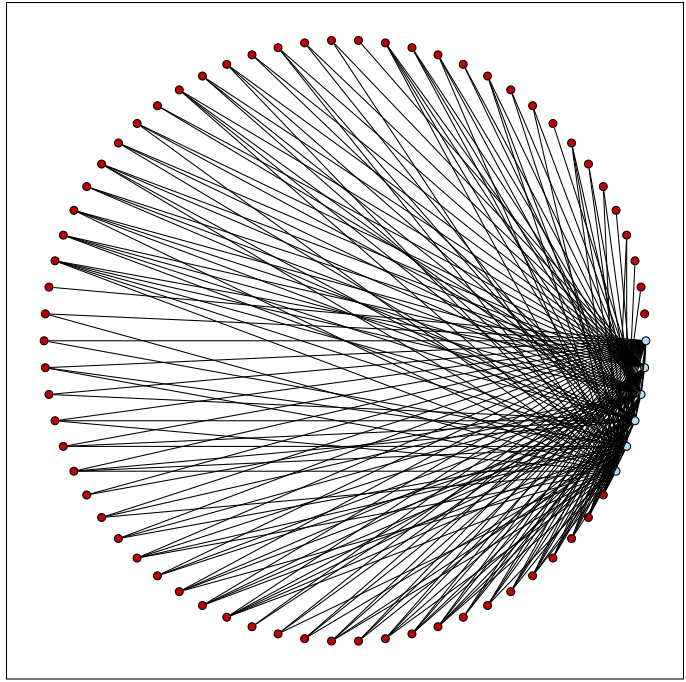
\begin{figure*}[htp]
\begin{center}

\begin{tikzpicture} 

\path[draw = black] (-4.5, -4.5) rectangle (4.5, 4.5); 

\def\distanceFromCenter{4}

\foreach \i in {1,2,3,4,5,6} {
	\path (\i*-36/7+36/7:\distanceFromCenter) node[blue node] (b\i) {};
}

\foreach \redindex/\redsubscript in {1/000000, 2/000001, 3/000010, 4/000011, 5/000100, 6/000101, 7/000110, 8/000111, 9/001000, 10/001001, 11/001010, 12/001011, 13/001100, 14/001101, 15/001110, 16/001111, 17/010000, 18/010001, 19/010010, 20/010011, 21/010100, 22/010101, 23/010110, 24/010111, 25/011000, 26/011001, 27/011010, 28/011011, 29/011100, 30/011101, 31/011110, 32/011111} {
	\path (\redindex*36/7:\distanceFromCenter) node[red node] (r\redsubscript) {};
}

\foreach \redindex/\redsubscript in {1/100000, 2/100001, 3/100010, 4/100011, 5/100100, 6/100101, 7/100110, 8/100111, 9/101000, 10/101001, 11/101010, 12/101011, 13/101100, 14/101101, 15/101110, 16/101111, 17/110000, 18/110001, 19/110010, 20/110011, 21/110100, 22/110101, 23/110110, 24/110111, 25/111000, 26/111001, 27/111010, 28/111011, 29/111100, 30/111101, 31/111110, 32/111111} {
	\path (\redindex*36/7+32*36/7:\distanceFromCenter) node[red node] (r\redsubscript) {};
}

\draw[edge] (b1) -- (b2) -- (b3) -- (b4) -- (b1) -- (b3);
\draw[edge] (b2) -- (b4);


\foreach \s in {0, 1} {
	\foreach \t in {0, 1} {
		\foreach \u in {0, 1} {
			\foreach \v in {0, 1} {
				\foreach \w in {0, 1} {
					\foreach \x in {0, 1} {
						\ifthenelse{1 > \s} {
						}
						{
							\draw[edge] (r\s\t\u\v\w\x) -- (b1);
						}
						\ifthenelse{1 > \t} {
						}
						{
							\draw[edge] (r\s\t\u\v\w\x) -- (b2);
						}
						\ifthenelse{1 > \u} {
						}
						{
							\draw[edge] (r\s\t\u\v\w\x) -- (b3);
						}
						\ifthenelse{1 > \v} {
						}
						{
							\draw[edge] (r\s\t\u\v\w\x) -- (b4);
						}
						\ifthenelse{1 > \w} {
						}
						{
							\draw[edge] (r\s\t\u\v\w\x) -- (b5);
						}
						\ifthenelse{1 > \x} {
						}
						{
							\draw[edge] (r\s\t\u\v\w\x) -- (b6);
						}
					}
				}
			}
		}	
	}
}
\end{tikzpicture}
\end{center}
\caption{
A drawing of $G'_2$, i.e., $CE(6)$, whose vertex set
consists of a clique (light blue) of six vertices
and an independent set (dark red) of 64 vertices with distinct neighborhoods.
} 
\label{fig:CE6full}
\end{figure*}
\end{center}
\begin{proof}
While the graph $CE(6)$ is defined unambiguously by the theorem statement,
we give the following definition of the graph family $CE(k)$ in order to assign unique names to the vertices of $CE(6)$,
and to be able to refer to its induced subgraphs.
Denote by $[k]$ the set of integers $\set{1, 2, \ldots, k}$.
For $k \in \Zplus$,
let 
$B(k) = \set{b_1, b_2, \ldots, b_k}$
be a set of $k$ light blue vertices,
and let
$R(k) = \set{r_A \mid A \subseteq [k]}$
be a set of $2^k$ dark red vertices.
Let $CE(k)$ be the graph on
$B(k) \uplus R(k)$ in which
$B(k)$ is a clique, 
$R(k)$ is an independent set,
and there is an edge between $b_i \in B(k)$ and $r_A \in R(k)$
if and only if $i \in A$.

First we present lemmas regarding 1-obstacle representations of $CE(4)$
that will prove instrumental in 
showing that $CE(6)$ does not have a 1-obstacle representation.
We do this by exploiting the hereditary nature of the $CE$ family, that is,
whenever $k' < k$, 
copies of $CE(k')$ can be found as an induced subgraph of $CE(k)$
in a color-preserving fashion.

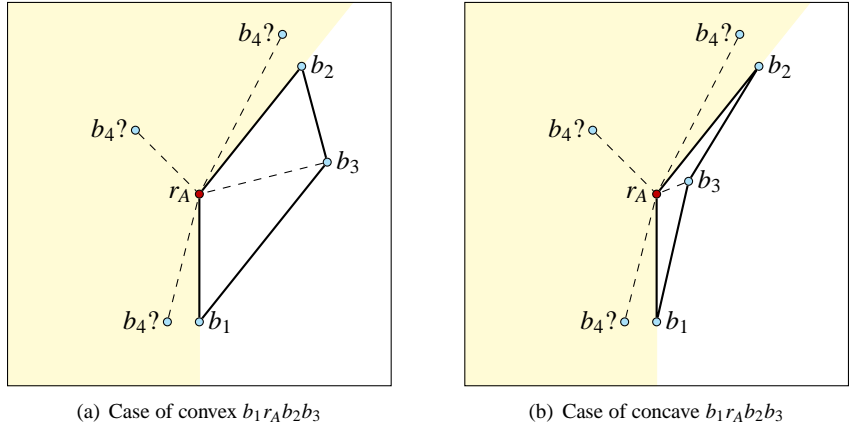
\begin{figure*}[htp]

\begin{center}
\subfigure[Case of convex $b_1 r_A b_2 b_3$]{
\begin{tikzpicture}[scale=0.85]
\newcommand{\figxmin}{-3}
\newcommand{\figxmax}{3}
\newcommand{\figymin}{-3}
\newcommand{\figymax}{3}
\draw[carve] (0,0) -- (2.4,3) -- (\figxmin, \figymax) -- (\figxmin, \figymin) -- (0, \figymin) -- (0,0);
\path (0, 0) node[red node] (rA) [label=left:$r_A$]{};
\path (0, -2) node[blue node] (b1) [label=right:$b_1$]{};
\path (1.6, 2) node[blue node] (b2) [label=right:$b_2$]{};
\path (2, 0.5) node[blue node] (b3) [label=right:$b_3$]{};
\draw[thick edge] (b1) -- (rA) -- (b2) -- (b3) -- (b1);
\draw[nonedge] (b3) -- (rA);
\path (-0.5, -2) node[blue node] (b4cand1) [label=left:$b_{4}?$]{};
\path (-1, 1) node[blue node] (b4cand2) [label=left:$b_{4}?$]{};
\path (1.3, 2.5) node[blue node] (b4cand3) [label=left:$b_{4}?$]{};
\draw[nonedge] (b4cand1) -- (rA);
\draw[nonedge] (b4cand2) -- (rA);
\draw[nonedge] (b4cand3) -- (rA);
\path[draw = black] (\figxmin, \figymin) rectangle (\figxmax, \figymax);
\end{tikzpicture}
\label{fig:quadConvex}
} 
\qquad
\subfigure[Case of concave $b_1 r_A b_2 b_3$]{
\begin{tikzpicture}[scale=0.85]
\newcommand{\figxmin}{-3}
\newcommand{\figxmax}{3}
\newcommand{\figymin}{-3}
\newcommand{\figymax}{3}
\draw[carve] (0,0) -- (2.4,3) -- (\figxmin, \figymax) -- (\figxmin, \figymin) -- (0, \figymin) -- (0,0);
\path (0, 0) node[red node] (rA) [label=left:$r_A$]{};
\path (0, -2) node[blue node] (b1) [label=right:$b_1$]{};
\path (1.6, 2) node[blue node] (b2) [label=right:$b_2$]{};
\path (0.5, 0.2) node[blue node] (b3) [label=right:$b_3$]{};
\draw[thick edge] (b1) -- (rA) -- (b2) -- (b3) -- (b1);
\draw[nonedge] (b3) -- (rA);
\path (-0.5, -2) node[blue node] (b4cand1) [label=left:$b_{4}?$]{};
\path (-1, 1) node[blue node] (b4cand2) [label=left:$b_{4}?$]{};
\path (1.3, 2.5) node[blue node] (b4cand3) [label=left:$b_{4}?$]{};
\draw[nonedge] (b4cand1) -- (rA);
\draw[nonedge] (b4cand2) -- (rA);
\draw[nonedge] (b4cand3) -- (rA);
\path[draw = black] (\figxmin, \figymin) rectangle (\figxmax, \figymax);
\end{tikzpicture}
\label{fig:quadConcave}
} 

\end{center}
\caption{
For the proof of Lemma \ref{whyDangerous}.
The red vertex $r_A$ is 
\dangerous{A} with $1,2 \in A$ and $3,4 \in \comp{A}$.
Without loss of generality,
$b_3$ is in $\ch{\angle{b_1 r_A b_2}}$ (unshaded)
while
$b_4$ is in the complement of $\ch{\angle{b_1 r_A b_2}}$ (shaded).
} 
\label{fig:dangerous}
\end{figure*}


We first establish some properties in 1-obstacle representations of small $CE$ graphs.

When considering obstacle representations for $CE(k)$,
for a fixed index set $A \subseteq [k]$
we 
denote
$[k] \setminus A$
by $\comp{A}$.
For a fixed placement of the blue vertices $B(k)$ in general position,
we say a point $p$ in general position with respect to $B(k)$ is
\dangerous{A} if there are distinct
$i_1, i_2 \in A$ and distinct
$i_3, i_4 \in \comp{A}$
such that
$\angle b_{i_1} p b_{i_2}$ separates $b_{i_3}$ from $b_{i_4}$.

\begin{lemma}
\label{whyDangerous}
For every integer $k \geq 4$,
every obstacle representation of $CE(k)$ 
in which some $r_A \in R(k)$ is \dangerous{A}
involves at least two obstacles.
\end{lemma}
\begin{proof}
For an arbitrary $k \geq 4$,
consider an obstacle representation of $CE(k)$ 
in which for a certain $A \subseteq [k]$, 
$r_A$ is \dangerous{A}.
Without loss of generality $1, 2 \in A$ and $3, 4 \in \comp{A}$
with $b_{3} \in \ch{\angle b_{1} r_{A} b_{2}}$
and $b_{4} \notin \ch{\angle b_{1} r_{A} b_{2}}$.
(See Fig.~\ref{fig:dangerous}.)

Then the quadrilateral $Q = b_{1} r_{A} b_{2} b_{3}$ is 
non-self-intersecting and has \pair{r_{A} b_{3}} as an internal diagonal.
Hence, an obstacle is needed inside $Q$,
which is interior-disjoint from the complement of $\ch{\angle b_{1} r_{A} b_{2}}$,
in order to block 
$\pair{r_{A} b_{3}}$.
Since $r_4 \notin \ch{\angle b_{1} r_{A} b_{2}}$,
so is \pair{r_{A} b_{4}}, therefore
a different obstacle must block 
$\pair{r_{A} b_{4}}$.
\end{proof}

To simplify the notation for red vertices,
from now on we will write the subscript
$i$ instead of $\set{i}$, and
$\comp{i}$ instead of $[k] \setminus \set{i}$
whenever convenient.
We will also write $B$ instead of $B(k)$ where the value of $k$ is clear from context.

\begin{lemma}
\label{CEkBluesConvex}
For every integer $k \geq 4$,
in every \mbox{1-obstacle} representation of $CE(k)$,
$B$ is in convex position.
\end{lemma}

\begin{figure*}[htp]
\begin{center}
\subfigure[Subcase of $r_\set{3,4}$ above $\seg{b_1 r_\set{1,4}}$]{
\begin{tikzpicture}[scale=1]
\path (-2, -1) node[blue node] (b1) [label=left:$b_1$]{};
\path (2, -1) node[blue node] (b3) [label=right:$b_3$]{};
\path (0, 2.6) node[blue node] (b2) [label=right:$b_2$]{};
\path (0, 2) node[blue node] (b4) [label=right:$b_4$]{};
\path (-0.5, 0.7) node[red node] (r34) [label=below:$r_\set{3,4}$]{};
\path (0.5, 0.8) node[red node] (r14) [label=below:$r_\set{1,4}$]{};
\draw [thick edge] (b1) -- (b2) -- (b3) -- (b4) -- (b1) -- (b3);
\draw [thick edge] (b2) -- (b4);
\draw [thick edge] (b1) -- (r14) -- (b4);
\draw [thick edge] (b3) -- (r34) -- (b4);
\draw [nonedge] (b1) -- (r34);
\draw [nonedge] (b3) -- (r14);
\path[draw = black] (-2.6, -1.3) rectangle (2.6, 3);
\end{tikzpicture}
\label{fig:evenconf}
} 
\qquad
\subfigure[Subcase of $r_\set{3,4}$ below $\seg{b_1 r_\set{1,4}}$]{
\begin{tikzpicture}[scale=1]
\path (-2, -1) node[blue node] (b1) [label=left:$b_1$]{};
\path (2, -1) node[blue node] (b3) [label=right:$b_3$]{};
\path (0, 2.6) node[blue node] (b2) [label=right:$b_2$]{};
\path (0, 2) node[blue node] (b4) [label=right:$b_4$]{};
\path (-0.5, -0.2) node[red node] (r34) [label=below:$r_\set{3,4}$]{};
\path (0.5, 0.8) node[red node] (r14) [label=below:$r_\set{1,4}$]{};
\draw [thick edge] (b1) -- (b2) -- (b3) -- (b4) -- (b1) -- (b3);
\draw [thick edge] (b2) -- (b4);
\draw [thick edge] (b1) -- (r14) -- (b4);
\draw [thick edge] (b3) -- (r34) -- (b4);
\draw [nonedge] (b1) -- (r34);
\draw [nonedge] (b3) -- (r14);
\path[draw = black] (-2.6, -1.3) rectangle (2.6, 3);
\end{tikzpicture}
\label{fig:unevenconf}
} 
\end{center}
\caption{
For the proof of Lemma \ref{CEkBluesConvex}, Case 1.
} 
\label{fig:r34andr14contr}
\end{figure*}
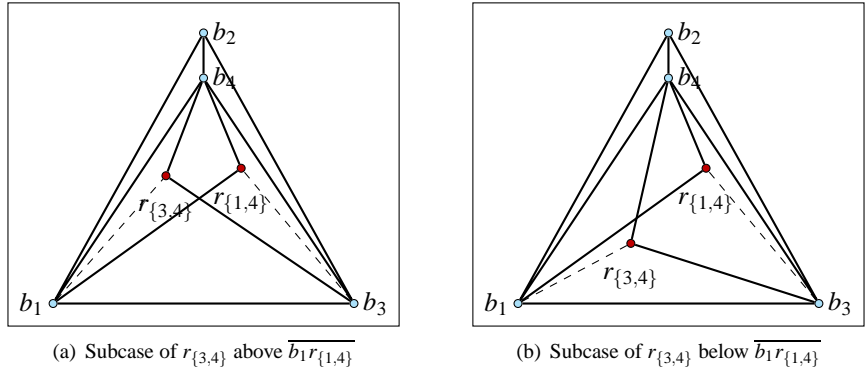

\begin{proof}
By Carath{\'e}odory's Theorem, it is sufficient to prove the result for $k=4$.

Assume for contradiction that 
we are given a \mbox{1-obstacle} representation of $CE(4)$ 
in which $B$ is not in convex position.
Without loss of generality,
$b_4$ is inside the triangle
$\tri b_1 b_2 b_3$.
There are two cases to consider.

\emph{Case 1:}
The obstacle is in \ch{B}.
Without loss of generality, 
the obstacle is inside $\tri b_1 b_4 b_3$.
The vertex $r_\set{1,4}$ has non-edges to $b_2$ and $b_3$,
so if it were outside of $\tri b_1 b_2 b_3$ then at least
one of these two non-edges would be outside of $\tri b_1 b_2 b_3$,
requiring a second obstacle.
Nor can $r_\set{1,4}$ be inside $\tri b_1 b_2 b_4$ or $\tri b_2 b_3 b_4$,
since that would cause its non-edge with $b_2$ to be in
inside that triangle,
again requiring a second obstacle.
A symmetric argument applies to $r_\set{3,4}$.
Notice that $r_\set{1,4} \in \ch{\angle b_4 b_2 b_3}$,
lest it be \dangerous{\set{1,4}}.
Likewise, $r_\set{3,4} \in \ch{\angle b_1 b_2 b_4}$,
lest it be \dangerous{\set{3,4}}.
Then without loss of generality, 
$r_\set{1,4}$ is inside
$\tri r_{\set{3,4}} b_4 b_3$
which $\pair{b_1 r_\set{3,4}}$ is outside of.
(See Fig.~\ref{fig:r34andr14contr}.)
Hence, distinct obstacles are required to block 
$\pair{b_1 r_\set{3,4}}$ and 
$\pair{b_3 r_\set{1,4}}$, a contradiction.

\emph{Case 2:}
The obstacle is outside of \ch{B}.
Then $r_\comp{4} \notin \ch{B}$ and
without loss of generality, $r_\comp{4} \in \ch{\angle b_1 b_4 b_3}$.
Hence the obstacle is inside $\tri b_1 b_3 r_\comp{4}$.
Since the quadrilateral $Q = b_1 b_4 b_3 r_\comp{4}$ is convex,
every point outside of $Q$ has a segment joining it to
$b_4$ or $r_\comp{4}$ without crossing $Q$.
Therefore, every remaining red vertex without an edge to $b_4$,
in particular, $r_\set{1,3}$,
is inside $Q$.
The introduction of $r_\set{1,3}$ into the drawing
results in interior-disjoint quadrilaterals
$Q' = b_1 r_\set{1,3} b_3 r_\comp{4}$
and
$Q'' = b_1 r_\set{1,3} b_3  b_4$.
(See Fig.~\ref{fig:CEkBluesConvexCase2}.)
Since 
$\pair{r_\set{1,3} r_\comp{4}}$
is inside
$Q'$
and 
$\pair{r_\set{1, 3} b_4}$ is inside $Q''$,
distinct obstacles are required to block these non-edges, a contradiction.
\end{proof}

\begin{figure*}[htp]
\begin{center}
\begin{tikzpicture}[scale=1.3]
\path (-1, 0) node[blue node] (b1) [label=left:$b_1$]{};
\path (1, 0) node[blue node] (b3) [label=right:$b_3$]{};
\path (0, 1.5) node[blue node] (b2) [label=right:$b_2$]{};
\path (0, 0.8) node[blue node] (b4) [label=right:$b_4$]{};
\path (0.7, -1.5) node[red node] (r4bar) [label=below:$r_\comp{4}$]{};
\path (0.2, -0.4) node[red node] (r13) [label=left:$r_\set{1,3}$]{};
\draw [thick edge] (b1) -- (b2) -- (b3) -- (b4) -- (b1) -- (b3);
\draw [thick edge] (b2) -- (b4);
\draw [thick edge] (b1) -- (r4bar) -- (b3);
\draw [thick edge] (b1) -- (r13) -- (b3);
\draw [nonedge] (b4) -- (r13);
\draw [nonedge] (b4) -- (r4bar);
\draw [nonedge] (r4bar) -- (r13);
\path[draw = black] (-2.2, -2) rectangle (2.2, 1.9);
\end{tikzpicture}
\end{center}
\caption{
For the proof of Lemma \ref{CEkBluesConvex}, Case 2.
The assumptions lead without loss of generality to
the configuration shown here, with
\pair{r_\set{1,3} b_4} and 
\pair{r_\set{1,3} r_\comp{4}} requiring distinct obstacles.
} 
\label{fig:CEkBluesConvexCase2}
\end{figure*}
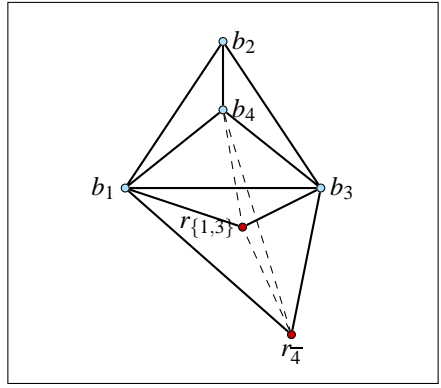

Now that we have some restrictions on the relative positions of blue vertices
in all \mbox{1-obstacle} representations of $CE(k)$ for all $k \geq 4$,
we pursue the question of where the red vertices can be positioned with respect to the blue vertices.

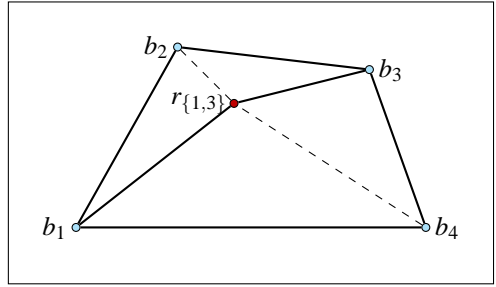
\begin{figure*}[htp]
\begin{center}
\begin{tikzpicture}[scale=1.5]
\path (0.6, 0.9) node[blue node] (b1) [label=left:$b_1$]{};
\path (1.5, 2.5) node[blue node] (b2) [label=left:$b_2$]{};
\path (3.2, 2.3) node[blue node] (b3) [label=right:$b_3$]{};
\path (3.7, 0.9) node[blue node] (b4) [label=right:$b_4$]{};
\path (2, 2) node[red node] (r13) [label=left:$r_\set{1,3}$]{};
\draw [thick edge] (b1) -- (b2) -- (b3) -- (b4) -- (b1);
\draw [thick edge] (b1) -- (r13) -- (b3);
\draw [nonedge] (b4) -- (r13) -- (b2);
\path[draw = black] (0, 0.4) rectangle (4.3, 2.9);
\end{tikzpicture}
\end{center}
\caption{
For the proof of Lemma \ref{mostRedsExteriorToBForCEk}, Case 2.
The vertex $r_\set{1,3}$ is \dangerous{\set{1,3}}.
} 
\label{fig:obstacleExteriorCE4}
\end{figure*}

\begin{lemma}
\label{mostRedsExteriorToBForCEk}
For every integer $k \geq 4$,
in every \mbox{1-obstacle} representation of $CE(k)$
the obstacle is outside of \ch{B},
and hence $R \setminus \set{r_B}$ is outside of \ch{B}.
\end{lemma}
\begin{proof}
It is clearly enough to establish the lemma in the special case $k=4$.

Assume for contradiction that 
we are given a \mbox{1-obstacle} representation of $CE(4)$ such that
the obstacle is in \ch{B}.
By Lemma \ref{CEkBluesConvex},
$B$ is in convex position.
Without loss of generality, 
$b_1 b_2 b_3 b_4$ is a clockwise enumeration of $B$.

\emph{Case 1:}
$r_\set{1,3} \notin \ch{B}$.
Imagine 
the polygon $b_1 b_2 b_3 b_4$ bounding 
\ch{B} as opaque:
Since it is convex, $r_\set{1,3}$ 
sees some side of \ch{B} in its entirety,
hence $r_\set{1,3}$ sees $b_i$ for a certain even $i$.
This means $\pair{r_\set{1,3} b_i}$ is outside \ch{B},
hence it will require a separate obstacle,
a contradiction.

\emph{Case 2:}
$r_\set{1,3} \in \ch{B}$.
(See Fig.~\ref{fig:obstacleExteriorCE4}.)
By the convexity of $B$,
$r_\set{1, 3}$
is \dangerous{\set{1, 3}}, a contradiction.
\end{proof}

We introduce some further terminology 
to use in the context of $CE(k)$ (for any integer $k > 0$)
for a fixed arrangement of $B$.
The following definitions are meant only for points outside of \ch{B}
and in general position with respect to $B$.

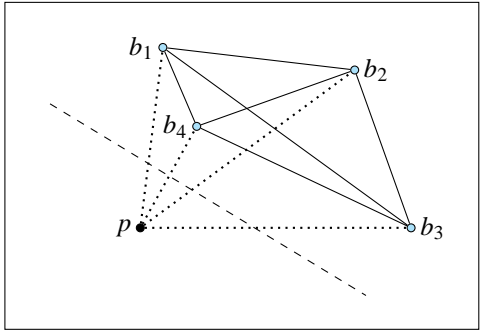
\begin{figure*}[htp]
\begin{center}
\begin{tikzpicture}[scale=1.5]

\tikzstyle{pToBlue} = [draw,thick,black,dotted]
\tikzstyle{sepLine} =  [draw,-,black,dashed]

\path[draw = black] (0.1, 0) rectangle (4.3, 2.9);

\path (1.5, 2.5) node[blue node] (b1) [label=left:$b_1$]{};
\path (3.2, 2.3) node[blue node] (b2) [label=right:$b_2$]{};
\path (3.7, 0.9) node[blue node] (b3) [label=right:$b_3$]{};
\path (1.8, 1.8) node[blue node] (b4) [label=left:$b_4$]{};

\path (1.3, 0.9) node (p) [label=left:$p$]{};

\draw [edge] (b1) -- (b2) -- (b3) -- (b4) -- (b1) -- (b3);
\draw [edge] (b2) -- (b4);

\draw [pToBlue] (p) -- (b1);
\draw [pToBlue] (p) -- (b2);
\draw [pToBlue] (p) -- (b3);
\draw [pToBlue] (p) -- (b4);
\draw [sepLine] (0.5, 2) -- (3.3, 0.3);

\end{tikzpicture}
\end{center}
\caption{
Illustration for the concepts \nice{A}, \polite{A}, and \flanked{A}.
For $CE(4)$ or $CC(4)$,
consider the given placement of $B$.
The point $p$ is 
{\nice{\set{2, 3}}},
{\polite{\set{4}}}, and
{\flanked{\set{1, 4, 3}}}.
} 
\label{fig:NicePoliteFlankedDangerous}
\end{figure*}

For a given $A \subseteq [k]$, let $B_A$ denote $\set{b_i \mid i \in A}$.
We say that a point $p$
is \nice{A} 
if some line through $p$
separates $B_A$ and $B_{\comp{A}}$ 
(vacuously true if $A \in\set{\emptyset, [k]}$).
We say that a point $p$ is
\polite{A}
if it sees $B_{A} \neq \emptyset$ between two non-empty parts of 
$B_\comp{A}$ that comprise $B_\comp{A}$.
If a point is \polite{\comp{A}},
we say it is \flanked{A}.
Observe that if $r_A$ is \flanked{A}, then an obstacle is required in a bounded face,
but not necessarily in the unbounded face.
Note that for every $A \subseteq [k]$,
every point $p \notin \ch{B}$ in general position with respect to $B$
is either \nice{A}, \polite{A}, \flanked{A}, or \dangerous{A}.
(See Fig.~\ref{fig:NicePoliteFlankedDangerous}.)

Now we can finish proving with relative ease
that $CE(6)$ does not admit a 1-obstacle representation.
Assume for contradiction that
we are given a \mbox{1-obstacle} representation of $CE(6)$.
By Lemma \ref{CEkBluesConvex}, $B$ is in convex position.
Without loss of generality,
$b_1 b_2 b_3 b_4 b_5 b_6$
is a clockwise enumeration of $B$.
By Lemma \ref{mostRedsExteriorToBForCEk},
$R \setminus \set{r_{B}}$ is outside of \ch{B}.
In particular, $r_\set{1,3,5}$ is outside of \ch{B}.
We will show that
every point outside of \ch{B} and in general position with respect to $B$ is
\dangerous{\set{1,3,5}}
by showing that it is
neither
\nice{\set{1,3,5}}
nor
\polite{\set{1,3,5}}
nor
\flanked{\set{1,3,5}}.

Clearly, no point
is \nice{\set{1,3,5}},
since $\set{b_1, b_3, b_5}$ is not linearly separable from 
$\set{b_2, b_4, b_6}$.

Assume for contradiction that
some point $p$ is \polite{\set{1,3,5}}.
Hence $p$ sees odd-subscripted blue vertices together
between two sets of even-subscripted blue vertices.
Then $p$ is \nice{\set{i, j}} for some $\set{i, j} \subseteq \set{2,4,6}$.
But $\pair{b_i b_j}$ 
is a diagonal of the bounding hexagon of $B$, 
which contradicts that it is linearly separable from $B \setminus \set{b_i, b_j}$.
By a symmetric argument, no point is \polite{\set{2,4,6}} (i.e., \flanked{\set{1,3,5}}) either.
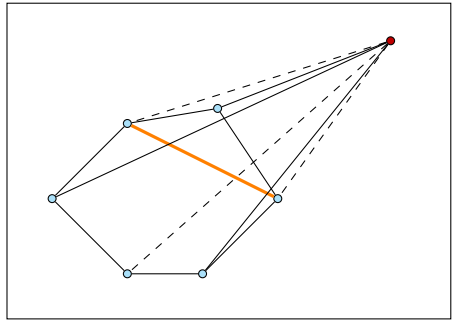
\begin{figure*}[htp]
\begin{center}
\begin{tikzpicture}[scale=1]
\tikzstyle{emphedge} = [draw,very thick,orange]	

\path[draw = black] (-2.6,-1.6) rectangle (3.3, 2.6);
\newcommand{\imax}{8}
\newcommand{\jmax}{\imax}

\path (-2, 0) node[blue node] (b1) [label=left:$$]{};
\path (-1, 1) node[blue node] (b2) [label=left:$$]{};
\path (0.2, 1.2) node[blue node] (b3) [label=left:$$]{};
\path (1, 0) node[blue node] (b4) [label=left:$$]{};
\path (0, -1) node[blue node] (b5) [label=left:$$]{};
\path (-1, -1) node[blue node] (b6) [label=left:$$]{};
	\draw[edge] (b1) -- (b2) -- (b3) -- (b4) -- (b5) -- (b6) -- (b1);

\newcommand{\mycirclesize}{8pt};

	\draw[emphedge] (b2) -- (b4);
	\path (2.5, 2.1) node[red node] (r) [label=left:$$]{};
	\draw[edge] (r) -- (b1);
	\draw[edge] (r) -- (b3);
	\draw[edge] (r) -- (b5);
	\draw[nonedge] (r) -- (b2);
	\draw[nonedge] (r) -- (b4);
	\draw[nonedge] (r) -- (b6);
\end{tikzpicture}

\end{center}
\caption{
For the proof of Theorem \ref{CE6HasNoOneObstacleRep}.
A red vertex $r_A$ adjacent exactly to blue vertices non-adjacent in the bounding polygon of $B$
is \dangerous{A} no matter what, as in this example.
} 
\label{fig:CE6bad}
\end{figure*}

Therefore, $r_\set{1,3,5}$ is
\dangerous{\set{1,3,5}},
requiring two obstacles, a contradiction.

Therefore, $G'_2$, i.e., $CE(6)$, has obstacle number greater than \emph{one}.
\end{proof}

\section{A 10-vertex $(2,2)$-colorable graph without a 1-obstacle representation}
\label{sec:CCNotAllAtMost1}
We showed in \cite{PS10_GraphsCombin} that a $(2,2)$-colorable 20-vertex graph $G_3$ has obstacle number greater than 1.
One can obtain $G_3$ from $CE(4)$ by adding all possible edges among the vertices in the independent set of 16 red vertices.
Here, we show that a 10-vertex induced subgraph of it, $G'_3$, also has obstacle number greater than 1.

Let $G'_3$ be the graph consisting of a clique of light blue vertices
$B = \set{b_i \mid i \in [4]}$, 
a clique of dark red vertices $R = \set{r_A \mid A  \in {[4] \choose 2}}$, 
and additional edges between every $b_i$ and every $r_A$ with $i \in A$.
(See Fig.~\ref{fig:CC}.)
\begin{center}
\begin{figure*}[htp]
\begin{center}
\begin{tikzpicture}[scale=0.6]
\path[draw = black] (-4.5, -5) rectangle (4.5, 5); 
\def\distanceFromCenter{4}
\path (90+1*120:\distanceFromCenter) node[blue node] (b1) [label=left:$b_1$] {};
\path (90+2*120:\distanceFromCenter) node[blue node] (b2) [label=right:$b_2$] {};
\path (90+3*120:\distanceFromCenter) node[blue node] (b3) [label=above:$b_3$] {};
\path (0, 0) node[blue node] (b4) [label=left:$b_4$]{};
\path (150+1*120:\distanceFromCenter) node[red node] (r12) [label=below:$r_{12}$] {};
\path (150+2*120:\distanceFromCenter) node[red node] (r23) [label=above:$r_{23}$] {};
\path (150+3*120:\distanceFromCenter) node[red node] (r13) [label=above:$r_{13}$] {};
\path (30+1*120:\distanceFromCenter*0.3) node[red node] (r14) [label=left:$r_{14}$] {};
\path (30+2*120:\distanceFromCenter*0.3) node[red node] (r24) [label=left:$r_{24}$] {};
\path (30+3*120:\distanceFromCenter*0.3) node[red node] (r34) [label=right:$r_{34}$] {};
\foreach \i in {1,2,3,4} {
	\foreach \j in {1,2,3,4} {
		\ifthenelse {\i < \j} {
			\draw[edge] (b\i) -- (r\i\j) -- (b\j) -- (b\i);
		} {}
	}
}
\foreach \i in {12,13,14,23,24,34} {
	\foreach \j in {12,13,14,23,24,34} {
		\ifthenelse {\i < \j} {
			\draw[edge] (r\i) -- (r\j);
		} {}
	}
}
\end{tikzpicture}
\end{center}
\caption{
A drawing of $G'_3$ with 3-fold rotational symmetry.
} 
\label{fig:CC}
\end{figure*}
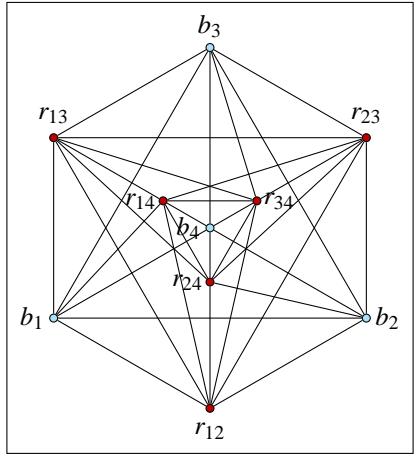
\end{center}

\begin{theorem}
\label{thm:CC_10vertex}
$G'_3$, a $(2,2)$-colorable graph on \emph{ten} vertices,
has obstacle number greater than \emph{one}.
\end{theorem}

\begin{proof}
We say that a polygon is \emph{solid} if all its edges are edges in $G'_3$. For three distinct points $p$, $q$, and $r$, we denote by $\angle pqr$ the union of the rays $\ray{qp}$ and $\ray{qr}$. For a point set $P$,
we denote by \ch{P} the convex hull of $P$ (the smallest convex set containing $P$).

Assume for contradiction that we are given a 1-obstacle representation of $G'_3$.
Following the terminology in the preceding section,
we shall say that a red vertex $r_A$ is \emph{\dang} if it is \dangerous{A}.
That is, a vertex $r_A$ is not {\dang} if and only if there are points $p$ and $q$ such that $\angle p r_A q$ strictly separates
$\set{b_i \mid i \in A}$ from the remaining blue vertices.
If some red vertex $r_A$ is {\dang}, then
\emph{two} obstacles will be required due to $\set{r_A} \cup B$, a contradiction.

\emph{Case 1:} 
$B$ is not in convex position. Without loss of generality,
$b_4$ is inside the triangle $\tri b_1 b_2 b_3$.

\emph{Subcase 1a:}
The obstacle is in \ch{B}. 
The proof of Lemma~\ref{CEkBluesConvex} Case 1
is based only on the vertices $b_1$, $b_2$, $b_3$, $b_4$, $r_\set{1,4}$ and $r_\set{2,4}$, under the same conditions,
hence that argument applies verbatim to yield a contradiction here.

\emph{Subcase 1b:}
The obstacle is outside of \ch{B}.
Let
$C_\set{1,2} = \ch{\angle b_2 b_4 b_1}$,
$C_\set{1,3} = \ch{\angle b_1 b_4 b_3}$, and
$C_\set{2,3} = \ch{\angle b_3 b_4 b_2}$.
Every red vertex is in precisely one of these regions and outside of \ch{B}.
Let $f: {[3] \choose 2} \to {[3] \choose 2}$ be the map such that
$r_A \in C_{f(A)}$ whenever $A \in {[3] \choose 2}$.
We will show that every possible assumption about $f$ leads to a contradiction.

Assume for contradiction that $f$ has a fixed point.
Without loss of generality, $r_\set{1,2} \in C_\set{1,2}$.
This means that $Q = b_2 b_4 b_1 r_\set{1,2}$ is a solid convex quadrilateral,
hence to block $b_4 r_\set{1,2}$,
the obstacle is inside $Q$. 
Then, $r_\set{3,4}$ must be inside $Q$ 
in order for the obstacle to block both $b_1 r_\set{3,4}$ and $b_2 r_\set{3,4}$.
But then, $\angle b_4 r_\set{3,4} r_\set{1,2}$ partitions $Q$
into disjoint quadrilateral regions with solid boundaries that
respectively contain $b_1 r_\set{3,4}$ and $b_2 r_\set{3,4}$.
Hence, two obstacles are required, a contradiction.
Therefore, $f$ has no fixed point.

Assume for contradiction that $f$ is not a permutation.
Without loss of generality,
$r_\set{1,3}$ and $r_\set{2,3}$ are both in $C_\set{1,2}$.
In order for both of these red vertices to not be {\dang},
$\lin{b_3 b_4}$ must separate 
$b_1 r_\set{1,3}$ and $b_2 r_\set{2,3}$.
Hence, $Q = b_2 b_1 r_\set{1,3} r_\set{2,3}$ is a solid, non-self-intersecting quadrilateral.
If $Q$ is concave, we get an immediate contradiction due to $Q$ 
separating its diagonals, both of which are non-edges in $G'_3$.
If $Q$ is convex, the obstacle is inside $Q$ in order to block its diagonals.
But since $r_\set{1,2}$ is outside of $C_\set{1,2}$, 
it does not meet \ch{Q}, requiring another obstacle, a contradiction.
Therefore, $f$ is a permutation.

Since $f$ is a permutation of three elements with no fixed point, it is cyclic.
Without loss of generality, $r_\set{1,2} \in C_\set{2,3}$ and $r_\set{1,3} \in C_\set{1,2}$.
In order to not be {\dang},
$r_\set{1,2}$ is on the same side of 
$\lin{b_1 b_4}$ as $b_2$,
and
$r_\set{1,3}$ is on the same side of
$\lin{b_3 b_4}$ as $b_1$.
These conditions ensure that $b_1 b_4$ does not meet $r_\set{1,2} r_\set{1,3}$.
If $b_2 b_4$ and $r_\set{1,2} r_\set{1,3}$ meet at some point $p$,
then 
the convex solid quadrilateral $b_1 r_\set{1,3} p b_4$
will have $b_2 r_\set{1,3}$ inside and $b_4 r_\set{1,2}$ outside, requiring
two obstacles, a contradiction.
If not, then $b_1 r_\set{1,3} r_\set{1,2} b_2 b_4$ is a non-self-intersecting 
solid pentagon with
$b_2 r_\set{1,3}$ inside and $b_4 r_\set{1,2}$ outside,
requiring
two obstacles, a contradiction.

Having exhausted all possibilities, 
we have shown that the assumptions of Subcase 1b lead to a contradiction.

\begin{figure*}[htp]
\begin{center}
\subfigure[Subcase 2a]{
\begin{tikzpicture}[scale=0.5]
\path[draw = black] (-1.5, -3.5) rectangle (9.5, 6); 
\node [red node, inner sep=0.75pt,label=below:$r_\set{1,3}$] (r13) at (0.5, -0.5) {};
\node [red node, inner sep=0.75pt,label=right:$r_\set{2,4}$] (r24) at (2.5, 5) {};
\node [blue node, inner sep=1pt,label=below:$b_3$] (3) at (2.5, 0.75) {};
\node [blue node, inner sep=1pt,label=right:$b_2$] (2) at (4,3) {};
\node [blue node, inner sep=1pt,label=right:$b_1$] (1) at (8,1) {};
\node [blue node, inner sep=1pt,label=below:$b_4$] (4) at (5,-2) {};
\draw [thick edge] (2) -- (3) -- (r13) -- (r24) -- (2);
\draw [thick nonedge] (3) -- (r24) -- (1);
\draw [edge] (2) -- (1) -- (4) -- (3) -- (r13) -- (r24);
\draw [edge] (2) -- (4);
\draw [edge] (1) -- (3);
\draw [edge] (r24) -- (4);
\draw [edge] (r13) -- (1);
\draw [nonedge] (2) -- (r13) -- (4);
\end{tikzpicture}
}
\qquad
\subfigure[Subcase 2b]{
\begin{tikzpicture}[scale=0.5]
\path[draw = black] (-1.5, -3.5) rectangle (9.5, 6); 
\node [red node, inner sep=0.75pt,label=left:$r_\set{2,4}$] (r24) at (3,2) {};
\node [red node, inner sep=0.75pt,label=right:$r_\set{1,3}$] (r13) at (1,5) {};
\node [blue node, inner sep=1pt,label=below:$b_3$] (3) at (2,-1) {};
\node [blue node, inner sep=1pt,label=below:$b_2$] (2) at (4,1) {};
\node [blue node, inner sep=1pt,label=right:$b_1$] (1) at (8,-0.3) {};
\node [blue node, inner sep=1pt,label=below:$b_4$] (4) at (5,-2) {};
\draw [thick edge] (3) -- (4) -- (1) -- (r13) -- (3);
\draw [thick edge]  (4) -- (r24) -- (r13);
\draw[thick nonedge] (1) -- (r24) -- (3);

\draw [edge] (3) -- (2) -- (1) -- (3);
\draw [edge] (r24) -- (2) -- (4);
\draw[nonedge] (4) -- (r13) -- (2);

\end{tikzpicture}
}
\caption{For the proof of Theorem~\ref{thm:CC_10vertex} Case 2.  The thick dashed non-edges require distinct obstacles.}
\label{fig:CCfig}
\end{center}
\end{figure*}
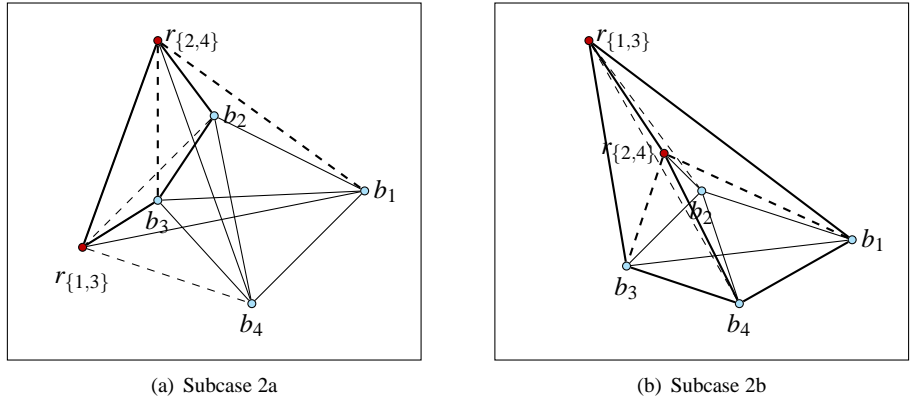

\emph{Case 2:}
$B$ is in convex position.
Without loss of generality, the bounding polygon of $B$ is
$b_1 b_2 b_3 b_4$.
To not be {\dang},

(i) $r_\set{1,3}$ and $r_\set{2,4}$ must lie outside of $\ch{B}$;

(ii) for $r_\set{1,3}$, either
$b_1, b_3 \in \ch{\angle b_2 r_\set{1,3} b_4}$ or
$b_2, b_4 \in \ch{\angle b_1 r_\set{1,3} b_3}$; and

(iii) for $r_\set{2,4}$, either
$b_1, b_3 \in \ch{\angle b_2 r_\set{2,4} b_4}$ or
$b_2, b_4 \in \ch{\angle b_1 r_\set{2,4} b_3}$.

\emph{Subcase 2a:}
$b_1, b_3 \in \ch{\angle b_2 r_\set{1,3} b_4}$
and
$b_2, b_4 \in \ch{\angle b_1 r_\set{2,4} b_3}$.
Without loss of generality, the quadrilateral $b_4 b_1 b_2 r_\set{1,3}$ is convex and has $b_3$ inside, and without loss of generality, the quadrilateral $b_3 b_4 b_1 r_\set{2,4}$ is convex and has $b_2$ inside. Hence, $b_2 b_3 r_\set{1,3} r_\set{2,4}$ is a solid convex quadrilateral with \pair{b_1 r_\set{2,4}} outside and \pair{b_3 r_\set{2,4}} inside. Therefore, two obstacles are required, a contradiction.

\emph{Subcase 2b:}
$b_2, b_4 \in \ch{\angle b_1 r_\set{1,3} b_3}$
or
$b_1, b_3 \in \ch{\angle b_2 r_\set{2,4} b_4}$.
Due to symmetry, we proceed assuming the former. Without loss of generality, $Q = b_3 b_4 b_1 r_\set{1,3}$ is a convex quadrilateral.
The obstacle is inside $Q$ due to $\pair{r_\set{1,3} b_4}$.
In order for $b_1 r_\set{2,4}$ and $b_3 r_\set{2,4}$ to be blocked,
$r_\set{2,4}$ is inside $Q$. Hence,
$\angle r_\set{1,3} r_\set{2,4} b_4$
partitions $\ch{Q}$ into two regions with solid boundaries that respectively contain
$\pair{b_1 r_\set{2,4}}$
and
$\pair{r_\set{2,4} b_3}$.
Therefore, two obstacles are required, a contradiction.

Therefore, $G'_3$ has obstacle number greater than \emph{one}.
\end{proof}

\bibliographystyle{amsplain}
\bibliography{obs}
\end{document}